\documentclass[twocolumn,usenatbib,amsmath,amssymb,amsbsy]{mn2e}

\usepackage{natbib}
\usepackage{amsbsy}
\usepackage{amssymb}
\usepackage{amsmath}
\usepackage{psfrag}
\usepackage{color}
\usepackage{graphicx}

\newcommand{\be}{\begin{equation}}
\newcommand{\ee}{\end{equation}}
\newcommand{\bear}{\begin{eqnarray}}
\newcommand{\eear}{\end{eqnarray}}

\newcommand{\bB}{{\bf B}}
\newcommand{\bE}{{\bf E}}
\newcommand{\bF}{{\bf F}}
\newcommand{\bJ}{{\bf J}}

\newcommand{\bv}{{\bf v}}

\newcommand{\re}{{\rm e}}
\newcommand{\rp}{{\rm p}}

\newcommand{\rx}{{\rm x}}

\newcommand{\z}{{\rm z}}

\newcommand{\rL}{{\rm L}}

\begin{document}

\title[Neutron star magnetospheres]
{The inside-out view on neutron star magnetospheres}

\author[Glampedakis, Lander \& Andersson]{K. Glampedakis$^{1,2}$,  S.K. Lander$^2$   \& N. Andersson$^{3}$ \\
  \\
  $^1$ Departamento de F\'isica, Universidad de Murcia, Murcia, E-30100, Spain \\
  $^2$ Theoretical Astrophysics, University of T\"ubingen, Auf der Morgenstelle 10, T\"ubingen D-72076, Germany \\
  $^3$ Mathematical Sciences and STAG Research Centre, University of Southampton, Southampton SO17 1BJ, UK}
 
\maketitle

\begin{abstract}

We construct hydromagnetic neutron star equilibria which allow for a non-zero electric current distribution in the exterior.
The novelty of our models is that the neutron star's interior field is in equilibrium with its magnetosphere, thus 
bridging the gap between previous work in this area which either solves for the interior assuming a vacuum exterior or
solves for the magnetosphere without modelling the star itself.   
We consider only non-rotating stars in this work, so our solutions are most immediately applicable to slowly-rotating systems such as magnetars. Nonetheless, we demonstrate that magnetospheres qualitatively resembling those expected for both magnetars and pulsars are possible within our framework. The ``inside-out'' approach taken in this paper should be more generally applicable to rotating neutron stars, where the interior and exterior regions are again not independent but evolve together.
\end{abstract} 
  


\section{Introduction}
\label{sec:intro}

Magnetic fields are ubiquitous in the world that surrounds us. They affect our everyday life and are key to many astrophysical phenomena as well. In fact, most of the information we have about the Universe has been gleaned from electromagnetic observations. Given this, it is natural that the origin, evolution and dynamics  of stellar magnetic fields remain important problems. 

As in many areas of physics, the extremes are particularly intriguing. Hence, it is natural that considerable attention has been given to the magnetars. These are observed as relatively young, slowly rotating neutron stars that show significant activity through bursts and occasional flares. Their phenomenology suggests that they have super-strong magnetic fields, of the order of $10^{15}\,\mbox{G}$~\citep{DT92}.
The challenge to understand the origin of these fields --- e.g. what kind of dynamo may act in the late stages of the core collapse 
when the neutron star is formed; their evolution, in the form of the coupled thermo-magnetic evolution in the star's solid crust; and their dynamics, say, the emission associated with giant flares --- is immense. As a result, the effort to understand the phenomenology of these systems has proceeded in steps where each problem is considered in isolation. This has progressed our understanding, but difficult issues remain unresolved. 

In this paper, we consider particular topics relating to a neutron star's magnetosphere. This is the region that surrounds a rotating magnetic star, where most of the electromagnetic emission is expected to originate. The motivation for studying this problem is obvious, and it is one with a long history. Most previous work focussed on the radio pulsar emission mechanism, a vexing problem that remains unsolved after more than forty years of observation. More recently, the nature of a magnetar's surroundings and the origin of gamma-ray flares and X-ray bursts have attracted significant attention. 

This present work attempts to contribute (in a relatively minor way) to both issues by modelling the magnetic field of a neutron  star in such a way that the interior is smoothly joined to the exterior. Existing models demonstrate a surprising dichotomy: they either solve the problem in the star's exterior without matching to an interior configuration, or model the interior problem assuming that the star is surrounded by vacuum. Both sets of solutions are obviously inconsistent. A star's magnetic field should be sourced by interior currents and the nature of the exterior magnetosphere must depend not only on the star's rotation and the exterior field strength but the interior field as well. The lack of consistent ``inside-out'' models becomes particularly problematic if one wants to understand to what extent the interior dynamics affects observed phenomena. Since this would involve a communication across the star's surface, the detailed physics in that region comes to the fore. Unfortunately, available models do not deal with this issue in a satisfactory manner. 

Taking  a small step towards more complete magnetar models, we extend recent work by \citet{lander09} in such a way that a localised magnetosphere is accounted for. In order to keep our task manageable in this first instance, we assume that the star is non-rotating. This is key because it allows us to avoid issues associated with the so-called light cylinder, a radius at which any matter corotating with the star would have to move at the speed of light, and where the problem in its usual incarnation becomes singular. It is also an obvious ``cheat'' because key phenomena associated with the transition from closed to open magnetic field lines (the pulsar emission mechanism, perhaps) cannot be accounted for. However, if we focus on the very slowly rotating magnetars then it stands to reason that the magnetospheric physics  near the star will be largely oblivious to any light-cylinder effects. 

We believe that our work provides an interesting complement to the model discussed by~\citet{BT07}. They suggest that a magnetar forms a magnetosphere through crust-cracking, which implants a ``twist'' (current) in the region immediately outside the star.  We present the first equilibrium models that allow for the presence of such currents, and demonstrate that one can, indeed, construct models with the features 
discussed by Beloborodov and Thompson. Notably, our modelling joins the interior and exterior currents smoothly. This means that surface currents are not required. This is an attractive feature of our model compared to other recent work (e.g. ~\citet{vigano11}, \citet{parfrey12}). In general, the introduction of a surface current seems somewhat arbitrary. More work is clearly needed to improve our understanding of the near-surface neutron star physics and establish whether one should expect surface currents to be present in a realistic model. Until such work is carried out, we feel that the introduction of arbitrary components in the model ought to be avoided. 


\section{State-of-the-art Magnetosphere construction}
\label{sec:art}

Even though our main interest will be in magnetars, it is important to understand how the  model connects with the modelling of normal radio pulsars. We will be making a number of simplifications in order to make progress on the construction of the twisted magnetosphere configuration that is expected to be relevant for magnetars, but in the future one would obviously like to remove these assumptions and address the complete problem. Hence, it is important to understand what the restrictions are and how our computational framework differs from the standard approach. 

Magnetospheres are thought to be composed of a tenuous magnetised electron-positron plasma with negligible density and pressure.
The plasma particles are assumed to be ``slaved'' to the electromagnetic field. This situation is the exact  opposite of that in the dense stellar interior, where the field is carried along with the fluid. 

Given the nature of the plasma in the exterior, it is not surprising that the so-called force-free approximation forms the backbone of most magnetosphere models~\citep{GJ69,Mestelbook}.
The essence of this approximation is the smallness of the various inertial, pressure and gravitational force terms compared to
 the electromagnetic Lorentz force $\bF_\rL$. The momentum equation describing the motion of particles of species $\rx$  (representing either electrons, $\rx=\re$, or `positrons',  $\rx=\rp$, these can be either actual positrons or protons) is written 
as
\be
q_\rx n_\rx \left ( \bE + \frac{\bv_\rx}{c} \times \bB \right ) = \mbox{inertia + pressure + gravity} \approx 0
\label{momeq1}
\ee
where $q_\rx,n_\rx$ and $\bv_\rx$ are the particle charge, number density and velocity respectively. 
Noting that each particle carries one unit of charge (positive/negative for positrons/electrons, obviously) we can sum the individual contributions from eqn.~(\ref{momeq1}) to get the force-free condition;
\be
\bF_\rL = \sigma_\re \bE + \frac{1}{c} \bJ \times \bB = 0
\label{FF}
\ee
where $\sigma_\re=e(n_\rp-n_\re)$ is the total charge density and $\bJ=e( n_\rp \bv_\rp- n_\re \bv_\re )$ is the total electric current (arising from the relative velocity of electrons and positrons). 

Combining  the pioneering corotating magnetosphere model of~\cite{GJ69} with the force-free assumption,  the additional restrictions of a stationary and 
axisymmetric system lead to the famous ``pulsar equation''. This is a second order, quasi-linear, elliptic equation that determines the 
magnetic field (\citet{sw73}, \citet{michel73}). In standard cylindrical coordinates $\{ \varpi,\varphi,z \}$, with the $z$-coordinate aligned with the system's symmetry axis, the pulsar equation takes the form
\be
\left (1- \frac{\varpi^2}{R_\rL^2} \right ) \Bigg [  \frac{\partial^2 u}{ \partial \varpi^2} +  \frac{\partial^2 u}{ \partial z^2} 
+ \frac{1}{\varpi}  \frac{\partial u}{ \partial \varpi} \Bigg ] 
- \frac{2}{\varpi} \frac{\partial u}{ \partial \varpi} = - f \frac{df}{du}
\label{pulsar0}
\ee
where $u$ is the stream function of the poloidal magnetic field and $f(u)$ is an unknown function representing
the toroidal magnetic field (see below for details). The equation has a singularity at the light cylinder radius, $R_\rL =c/\Omega$, 
where $\Omega$ is the angular rotation frequency of the star.

Decades of  effort have led to an established method for obtaining solutions to the pulsar equation, 
and thereby building neutron-star magnetospheres. The algorithm, which was developed by~\cite{CKF},  imposes  \emph{fixed} boundary conditions at the stellar surface -- usually that of a dipolar magnetic field -- and at infinity.
The algorithm then calculates $f(u)$ iteratively, making sure that the solution is well-behaved at the light cylinder.
This solution is self-consistent, in the sense that the particle velocities are kept well below the speed of light and
the force-free approximation is not violated. Subsequent work (e.g.~\citet{spitkovsky06}, \citet{kalapotharakos12}) 
has extended the method to more generic systems with non-axisymmetry and resistivity but the 
basic architecture of the approach is preserved; a fixed magnetic field is imposed at the stellar surface and
the global solution (through the calculation of $f(u)$) is regular across the light cylinder. In essence, these models treat 
the magnetosphere as a system that is completely detached from the stellar interior. This is the ``exterior'' approach to the problem. 

In our opinion, there is a fundamental problem with the exterior approach, relating to the matching at the star's surface. While in principle the approach of \citet{CKF} can be adjusted to match to any desired surface field, there is no way that the actual surface field will be known unless the interior problem is solved in conjunction with the exterior one. This point is fairly trivial, simply suggesting that it does not make much sense to glue together an interior solution relevant for an exterior vacuum to a force-free magnetosphere model, no matter how 
realistic the latter may be. In order to achieve true realism one would have to include the star itself in the system. 
This seems natural anyway given that the magnetic field, and hence the magnetosphere, is sourced by the currents in the star.  
The process of building a magnetosphere thus ought to entail the simultaneous solution of the pulsar equation \emph{and} of the magnetic field equilibrium in the stellar interior. 

We are not going to pretend that this is a simple problem. After all, if you are struggling to make progress on each of the two parts involved, then what chance have you got to figure out how to combine them? There are, however, key issues where we can make progress. We can, for example, try to improve our understanding of the nature of the transition from the star's core to its exterior. This involves resolving the issue of whether one should expect surface currents. Such currents are ``unattractive'' from the theory point of view as they involve a  degree of arbitrariness. If surface currents are, indeed, present then one would expect their nature to be determined by the physics. This fact has not been considered in any of the studies where such currents have been employed. In absence of a detailed model, we believe it makes sense to limit the amount of freedom in the model by excluding surface currents. Hence, our model guarantees a smooth transition from the interior to the exterior. All components of the magnetic field remain continuous.  

In developing the ``inside-out'' approach we are, at least at this initial stage, forced to make simplifications and approximations. Hence, we solve for the interior magnetic field together with the exterior magnetosphere in the non-rotating limit ($\Omega=0$). In other words, we solve the pulsar equation (\ref{pulsar0}) in the limit $R_\rL \to \infty$. This is convenient because it removes the singularity associated with the light-cylinder and therefore 
we do not need to consider the associated regularity conditions. 

How reasonable is the $R_\rL \to \infty$ approximation? The answer may depend on what we are actually interested in. For typical neutron star spin-periods the light cylinder is located several hundred 
stellar radii away from the surface (for magnetars this distance is about a factor of a hundred bigger) and therefore our model may provide
a good approximation of the  magnetosphere in a region extending several stellar radii from the surface. This part of the magnetosphere should be relatively oblivious to the physical conditions imposed by the light cylinder far away. On the other hand the model
is obviously not in any sense global and for a rotating star it must break down when $\varpi \gtrsim R_\rL$.

The upshot of this is that the  solutions we construct are more suitable for slowly spinning systems like
magnetars. Indeed,  magnetars have been \emph{exclusively} modelled as non-rotating both with regard to their dynamics
and the  structure of the magnetosphere (e.g. \citet{thompson02}, \cite{pavan09}, \citet{vigano11}). 
The association with magnetars is also promising because they are expected to have a ``twisted' magnetosphere'' \citep{thompson02} 
with a strong toroidal magnetic field in the  region of closed field lines near the star. As we will soon see, this property is closely 
captured by our model.


\section{Formalism}
\label{sec:formalism}

We aim to construct a simple model for a neutron star, taken to be a barotropic magnetised fluid ball, coupled to a magnetosphere, 
represented by a magnetised force-free plasma. As we have already discussed, we simplify the problem by requiring the solution to be both stationary and axisymmetric. 
Finally, we consider the problem in Newtonian gravity. Under these conditions, the hydromagnetic equilibrium in the stellar interior is described by the force balance between
pressure, gravity and magnetic field:
\be
\nabla p + \rho \nabla \Phi  = \bF_{\rm mag}
\label{euler1}
\ee
where $p$ is the pressure and $\Phi$ is the gravitational potential.
The constraint $\nabla \cdot \bB = 0$ implies that a magnetic field has just two degrees of freedom. Using a pair of scalar streamfunctions 
$u(\varpi,\z)$ and $f(\varpi,z)$ we may write our axisymmetric field in an automatically divergence-free fashion:
\be
\bB = \frac{1}{\varpi} \left [ \nabla u \times \hat{\varphi} + f \hat{\varphi}  \right  ]
\label{Bdeco}
\ee
The magnetic (Lorentz) force is given by
\be
\bF_{\rm mag} = \frac{1}{c}\, \bJ \times \bB =  \frac{1}{4\pi} (\nabla \times \bB ) \times \bB 
\label{Fmag}
\ee
where $\bJ$ is the total electric current. Taking the curl of \eqref{euler1} and using the barotropy
property of our model's matter, $p=p(\rho)$, we have
\be
\nabla \times \left ( \frac{1}{\rho} \bF_{\rm mag}  \right ) = 0
\label{curlF}
\ee
The imposed axisymmetry requires $F_{\rm mag}^\varphi =0$ which in turn leads to the functional
dependence $f=f(u)$. Then it is easy to show that 
\be
\bF_{\rm mag}=\rho\nabla M
\label{FyM}
\ee
where $M=M(u)$ is another scalar function (which obviously makes (\ref{curlF}) an identity).

Using (\ref{Bdeco}) in (\ref{FyM}) to calculate $\bF_{\rm mag}$ leads to the so-called
Grad-Shafranov equation~\citep{grad_rubin, shafranov}, which governs the hydromagnetic equilibrium in the stellar interior:
\be
\frac{\partial^2 u}{\partial \varpi^2}-\frac{1}{\varpi}\frac{\partial u}{\partial \varpi}
+\frac{\partial^2 u}{\partial z^2} = -4\pi\rho\varpi^2\frac{dM}{du} - f\frac{df}{du}
\label{GS}
\ee
In this equation, the  two functions $M(u)$ and $f(u)$ may be freely specified (modulo regularity and symmetry requirements). 
Through specific choices one may place physical restrictions on the equilibrium solutions; see e.g. \cite{ciolfi09} and \citet{lander12} 
for a discussion.

It is also informative to expand Amp\`ere's law in terms of the stream functions: 
\be
\nabla \times \bB= \frac{4\pi}{c}\bJ=\frac{df}{du} \bB + 4\pi\rho\varpi\frac{dM}{du} \hat{\varphi}
\ee
The first term on the right-hand side  describes the force-free part of the current while the second term represents a purely
azimuthal plasma flow.

Moving on, we next consider magnetic equilibrium in the star's exterior, i.e.  the magnetosphere. The simplest choice one can make is to 
assume that the star is surrounded by vacuum, removing the presence of  any charges or currents and effectively imposing $\nabla \times \bB =0$. 
This choice of an irrotational $\bB$ field is indeed commonplace in studies of hydromagnetic equilibria in the interior of neutron stars, 
see \citet{haskell08}, \citet{ciolfi09},\citet{lander12} for some recent examples. In essence, previous work on this subject has combined fairly advanced models for the neutron star interior with a rather primitive ``magnetosphere'' model.

As we have already discussed, our aim here is to calculate magnetic equilibria with an improved treatment of the magnetosphere.
In particular, we assume a non-vacuum magnetosphere filled with low-density plasma and where the force-free approximation, eqn~(\ref{FF}), is valid. 

It is worth noting that, as we ignore rotation, the electric field (and consequently the net charge density) is zero:
\be
\bE  = - \frac{1}{c} ( \varpi \Omega  \hat{\varphi} ) \times  \bB = 0
\ee
The magnetospheric current thus consists of particles sliding along the field lines, that is 
\be
\nabla \times \bB= \frac{4\pi}{c} \bJ = \frac{df}{du}\bB
\ee
and the force-free equation reduces to
\be
( \nabla \times \bB ) \times \bB = 0
\label{FL}
\ee
This identifies the exterior magnetic field as what is known as a \emph{Beltrami} vector field. 

Adopting the same ansatz (eqn.~(\ref{Bdeco})) as before we can again produce a Grad-Shafranov equation
\be
\frac{\partial^2 u}{ \partial \varpi^2} +  \frac{\partial^2 u}{ \partial z^2} - \frac{1}{\varpi}  \frac{\partial u}{ \partial \varpi} 
= - f \frac{df}{du}
\label{pulsar1}
\ee
Note that this equation coincides with the $R_\rL \to \infty$ limit of the pulsar equation~(\ref{pulsar0}). 
Compared to its counterpart in the stellar interior, Eqn. \eqref{pulsar1} displays the same freedom associated with the unspecified toroidal function $f(u)$
while lacking the degree of freedom associated with $M(u)$. The latter property follows from the fact
that eqn.~(\ref{curlF}) is weaker than the force-free condition (\ref{FL}).

We now see that the formalism provides us with a simple way to extend the interior solution to the magnetosphere. As is obvious from the two 
 Grad-Shafranov equations 
(\ref{GS}) and (\ref{pulsar1}), the exterior equation is simply the $\rho \to 0$ limit of the interior one (this limit is appropriate 
given that the magnetosphere is many orders of magnitude less dense than the stellar matter). Any given choice of functions $f(u)$ and
$M(u)$ leads to a consistent ``global'' calculation of the magnetic equilibrium without $\bB$-field discontinuities at the stellar surface.


\section{Numerical implementation}
\label{sec:numerics}

The formalism laid out in the previous Section provides us with a strategy for constructing non-rotating stellar models with non-trivial current-carrying magnetospheres. 
We need only solve the Grad-Shafranov equation \eqref{GS}
for the interior \emph{and} the exterior of the star. First we use the following vector identity for axisymmetric systems:
\be
\frac{\varpi}{\sin\phi}\nabla^2\left(\frac{u\sin\phi}{\varpi}\right)
 = \left( \frac{\partial^2}{\partial\varpi^2}-\frac{1}{\varpi}\frac{\partial}{\partial\varpi}
               +\frac{\partial^2}{\partial z^2} \right) u
\ee
to rewrite the Grad-Shafranov equation as a ``magnetic Poisson
equation'' involving the Laplace-type operator:
\be
\nabla^2\left(\frac{u\sin\phi}{\varpi}\right)
 = -\left(\frac{f}{\varpi}\frac{df}{du}+4\pi\varpi\rho \frac{dM}{du}\right)\sin\phi.
\ee
Together with this we need the usual equations governing equilibrium in a barotropic fluid star\footnote{We could, in principle, allow for composition-gradient stratification
of the stellar matter by using the scheme described in \citet{LAG}. For this study, however, 
our focus is on a more advanced model for the exterior.}. These consist of the Euler equation (\ref{euler1}) and
the Poisson equation
\be
\nabla^2\Phi = 4\pi G\rho,
\ee
supplemented by a polytropic equation of state
\be
p=p(\rho)=k\rho^2
\ee
where $k$ is a constant.

We solve this system of equations in integral form, making use of a non-linear numerical scheme~\citep{tomi_eri,lander09}
which iterates in $\rho$ and $u$, that is, we account for the effect of the pressure-density relation on the magnetic field
distribution, and the back-reaction of the field on the fluid.

We employ the usual Green's function to solve the two Poisson equations,
\be
G({\bf r,r'})=-\frac{1}{4\pi|{\bf r-r'}|},
\ee
which implicitly includes the correct behaviour at infinity
($\Phi=\mathcal{O}(r^{-1}),B=\mathcal{O}(r^{-3})$).

In constrast to other pulsar magnetosphere studies (e.g.~\citet{CKF}), we do not iterate directly in $f(u)$, but fix
its functional form at the outset. Similarly we fix $M(u)= \mbox{const} \times u$. The values of $f(u)$ and $M(u)$ across the 
system will, however, update as $u$ changes over iterative steps. Our approach makes sure that during the iteration any 
(probably unphysical) surface currents are avoided. 


\section{Results: magnetosphere solutions}
\label{sec:results}

\begin{figure}
\begin{center}
\begin{minipage}[c]{\linewidth}
\includegraphics[width=\textwidth]{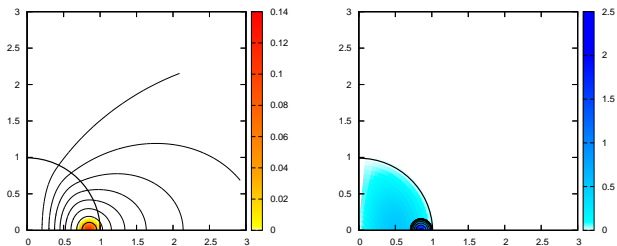}
\end{minipage}
\caption{\label{fig:vacuum} 
An example of magnetic field equilibrium for a neutron star model without exterior current. 
We show the direction of poloidal field/current with the lines and toroidal-field/current magnitude with the colour scales. The numerical domain is
expressed in units of the stellar radius $R$ -- in this and subsequent figures this domain is $0\leq r \leq 3 R$
where $r$ is the spherical coordinate radius.
Left panel: 
a typical twisted-torus magnetic field configuration, with $\zeta=0.1$ and with a vacuum exterior ($\nabla \times \bB=0$); 
no exterior current or toroidal field. The magnetic energy contained in the toroidal field component is $2.9\%$ 
of the total magnetic energy. Right panel: electric current distribution, $\bJ/c = \nabla \times \bB /4\pi$, for the magnetic equilibrium shown in the left-hand panel.} 
\end{center}
\end{figure}

The nature of axisymmetric and stationary magnetic equilibria depends heavily on the user-specified toroidal function $f(u)$. In calculations where the exterior is assumed
vacuum (see for instance, \citet{lander09,ciolfi09}) $f(u)$ was fitted inside the \emph{last closed} poloidal 
field line (thus ensuring the absence of exterior currents):
\be 
f(u) = \begin{cases}
                      a(u-u_{\rm int})^\zeta  & u> u_{\rm int}\\
                      0                          & u \leq u_{\rm int},
          \end{cases}
\label{fvac}
\ee
where $a$ and $\zeta$ are constant parameters and $u_{\rm int}$ is the value of the stream function associated with 
the last closed poloidal line. An example of this ``twisted torus'' equilibrium is shown in Fig.~\ref{fig:vacuum} (left panel)
for the specific choice $\zeta = 0.1$ (in each case $\zeta$ is chosen to give the largest possible percentage of toroidal field; 
the value of the amplitude $a$ sets the overall scale and is of less importance in the context of this work). The corresponding current distribution is shown in 
the right panel of Fig.~\ref{fig:vacuum} and it is easy to see that it is confined inside the star.

With our non-vacuum model it is possible to build more general (and more realistic) configurations. These are Beltrami-type magnetospheres and we will discuss several examples in the following sections.

\subsection{Magnetosphere with confined toroidal field}
\label{sec:confined}

The most straightforward extension of the vacuum model is to take the same functional form
$f(u)$ as above, but fitting the toroidal field to a larger contour of $u$, i.e.:
\be 
f(u) = \begin{cases}
                      a(u-\lambda u_{\rm int})^\zeta  & u>     \lambda u_{\rm int}\\
                      0                                   & u\leq \lambda u_{\rm int}.
       \end{cases}
\label{fconf}
\ee
The new parameter $\lambda < 1$ controls the size of the toroidal field region in the magnetosphere.

\begin{figure}
\begin{center}
\begin{minipage}[c]{\linewidth}
\includegraphics[width=\linewidth]{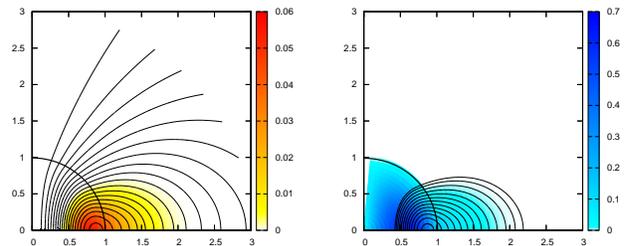}
\end{minipage}
\caption{\label{fig:confined}
Magnetic equilibrium allowing for exterior current. Left panel: magnetic field of a neutron star with a force-free magnetosphere, i.e. with parallel electric current and magnetic field ($\nabla\times\bB\ \|\ \bB$), in the equatorial lobe. In this
example $\zeta=0.5$. The magnetic energy contained in the toroidal field component is $7.1\%$ of the total.
Right panel: the corresponding electric current distribution $\bJ/c$
(as before we plot the direction of poloidal current (lines) and magnitude of toroidal current (colour code)). 
Note the presence of poloidal and toroidal currents in the magnetosphere.}
\end{center}
\end{figure}

A representative solution is shown in Fig.~\ref{fig:confined} for $\lambda=1/2$ and $\zeta=0.5$. The current distribution corresponding to
this magnetosphere solution is shown in the right panel of the same figure. As expected, the toroidal field in the exterior is sourced by a 
poloidal current flowing across the stellar surface. Solutions of this type have slightly more of the magnetic energy in the toroidal field component, compared with corresponding vacuum-exterior solutions, but they are still poloidal-dominated in a global sense. Locally however, in the environs of the magnetosphere, the toroidal component becomes dominant.

This type of equilibrium could be envisaged as a ``magnetar magnetosphere'', as it is qualitatively similar to the magnetar corona model
discussed by \citet{BT07}. One could imagine a magnetar with an initial field configuration like that of Fig.~\ref{fig:vacuum} suffering a crustquake, expelling poloidal current (toroidal field), then rearranging into an equilibrium solution like that shown in Fig.~\ref{fig:confined}.

\subsection{Magnetosphere with unconfined toroidal field}

\begin{figure}
\begin{center}
\begin{minipage}[c]{\linewidth}
\includegraphics[width=\linewidth]{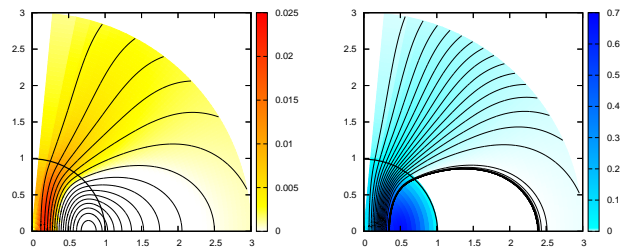}
\end{minipage}
\caption{\label{fig:unconfined}
Left panel: 
Magnetosphere solution with unconfined toroidal field, built according to the prescription in eqn.~(\ref{funconf}). 
The magnetic energy of the toroidal field (produced by integrating over the numerical domain) is only $1.5\%$ of the total magnetic
energy. For this example $\zeta=0.1$.
Right panel: 
the corresponding current distribution, depicted in the same way as in the previous figures.}
\end{center}
\end{figure}

An alternative type of magnetosphere can be built by fitting the toroidal field 
\emph{outside} a given poloidal field contour. An example of such solutions is provided
by
\be  
f(u) = \begin{cases}
       au(\lambda u_{\rm int}-u)^\zeta   & u<     \lambda u_{\rm int} \\
       0                                     & u\geq \lambda u_{\rm int}.
       \end{cases}
\label{funconf}
\ee
A bit more care is needed here though --- one must also ensure $f(u)=0$ along the $z$-axis, to avoid a divergent $B_\phi$ component. An equilibrium with the above form of $f$ is shown in Fig.~\ref{fig:unconfined}. The percentage of magnetic energy in the toroidal component is $1.5\%$, but this is an under-estimate as the integral is only over the numerical domain (and the toroidal component decays at a slower rate than the poloidal one). There is no limiting case of this `unconfined' solution which produces a vacuum-exterior model like Fig.~\ref{fig:vacuum} (except with $B_\phi\to 0$), in contrast with the confined-magnetosphere solution of the last subsection.

Within our framework, this unconfined type of solution may be regarded as a ``pulsar magnetosphere'', in the sense that the
toroidal field occupies the portion of the magnetosphere where the open field lines would have been located if the
neutron star were rotating. Of course, our $f(u)$ function is not adjusted for consistency with any light-cylinder boundary conditions; it is, however, adjusted for consistency with the star's interior.

\begin{figure}
\begin{center}
\begin{minipage}[c]{\linewidth}
\includegraphics[width=\linewidth]{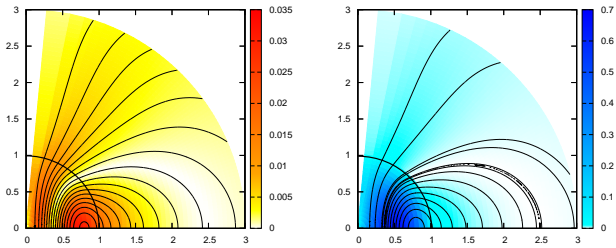}
\end{minipage}
\caption{\label{fig:double_magneto}
A ``mixed magnetosphere'', with toroidal field allowed to exist around the poles and 
in the near-equatorial region. For this particular solution $\zeta=0.2$ and $4.4\%$ of the magnetic energy
is in the toroidal component. Both the magnetic field and the current distribution are superpositions of
the previously discussed solutions.}
\end{center}
\end{figure}

Finally, the previous solutions can be combined to produce a mixed-type magnetosphere where the toroidal 
field ``lives'' around the pole \emph{and} in the field line region near the equator, see Fig.~\ref{fig:double_magneto}.
In this sense, this solution has a ``global'' toroidal field structure.

Simpler solutions with equally ``global'' toroidal fields can be produced by the previous models, by a suitable choice of the
contour line $u_{\rm int}$ (i.e. by pushing this boundary towards the axis or the equator). From the point of view of the exterior 
toroidal field structure, all these global solutions can be thought as being qualitatively similar to the self-similar twisted magnetosphere 
models of~\citet{thompson02} and of~\citet{pavan09}.

Other equilibrium solutions would be possible; the main limitations are making sure the toroidal field does not 
diverge at the pole, and whether or not the numerical scheme successfully iterates to a solution.


\subsection{Adding rotation}
\label{sec:rotation}

Although we have limited our analysis to  non-rotating systems it is fairly straightforward to 
approximate the effect of (weak) rotation on the magnetic equilibria. 
Specifically, we  solve the same system of equations as before, taking pre-specified exterior current distributions
through the function $f(u)$, but allowing for rigid rotation of the star. This gives us some streamfunction $u_0$, whose structure is somewhat affected by the now non-spherical surface shape of the star.
We then turn to the pulsar equation and assume a slow-rotation approximation in which the streamfunction in the light-cylinder term
is replaced by $u_0$, i.e. 
\be
\frac{\partial^2 u}{ \partial \varpi^2} +  \frac{\partial^2 u}{ \partial z^2} 
- \frac{1}{\varpi} \frac{\partial u}{ \partial \varpi} 
 = - f \frac{df}{du} +\frac{\varpi^2}{R_L^2}\nabla^2 u_0.
\ee
This equation is solved with iteration to produce the rotating model streamfunction $u$ --- without, however,
imposing any boundary condition at the location of the light cylinder (which anyway lies well outside our numerical domain).  

The magnetospheres produced through this exercise turn out to be very similar to those of the non-rotating system, even for a neutron star model rotating at 700 Hz, and are therefore not shown here.  They do, however, provide some justification for using a non-rotating ansatz, at least when studying the region close to the star (our numerical domain is $0\leq r\leq 3R$).


\section{Conclusions and discussion}
\label{sec:conclusions}

We have improved on previous models for magnetised neutron star equilibria by ensuring that the interior is smoothly joined to a more realistic exterior magnetosphere. As far as we are aware, this is the first serious step towards a global solution of this problem. Our models do not rely on surface currents or a strained crust; our pure-fluid model is equivalent to a neutron star with a relaxed crust. These two effects would tend to cause dissipation or rearrangement of the field, so by avoiding them we feel our equilibrium models may represent longer-lived magnetospheres.

Previous work on equilibrium solutions has either solved the problem in the star's exterior without matching to an interior configuration, or considered the interior problem assuming that the star is surrounded by vacuum. Neither set of models will lead to a realistic and consistent configuration. We argue that the ``inside-out'' approach is natural  since a star's magnetic field should be sourced by interior currents and the nature of the exterior magnetosphere must depend not only on the star's rotation and the exterior field strength but the interior field as well. A more detailed solution to this problem requires a better understanding of the communication across the star's surface and the detailed physics in that region. Such work is urgently needed if we want to make progress on a number of topical issues. 

In order to keep the problem tractable in this first instance, we have assumed that the star is non-rotating. This swept problems associated with the so-called light cylinder under the carpet, but it also means that the model cannot support key phenomena associated with the transition from closed to open magnetic field lines. This may not be an urgent problem, as  long as we focus on the slowly rotating magnetars. However, in order to proceed towards a general system we need to move beyond non-rotating stars. 
Since this is an important issue, it makes sense to close the paper with a few comments on the nature of the problem. 

\begin{figure}
\begin{center}
\begin{minipage}[c]{\linewidth}
\includegraphics[width=\linewidth]{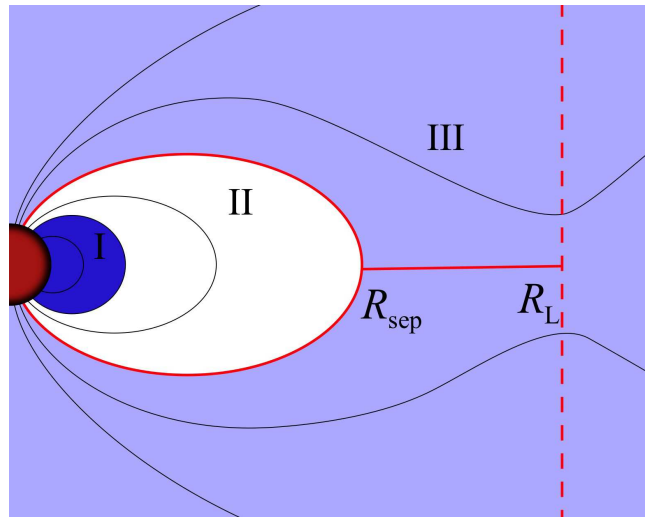}
\end{minipage}
\caption{\label{fig:cartoon}
In typical magnetosphere solutions, the separatrix between closed and open field lines is assumed to coincide with the light cylinder radius at the equator, hundreds or thousands of stellar radii from the star. If this is true, or approximately true, our zero-rotation assumption is a good approximation to the structure of the magnetosphere near a magnetar. This scenario is shown above: the equatorial lobe (I) is our magnetar magnetosphere and the white region surrounding it (II) is vacuum containing closed field lines. Our approximation breaks down beyond the separatrix, region III, where the field lines are expected to be open and the pulsar equation should be solved, but close to the star this is only a tiny region around the pole. If, for any reason, $R_{\rm sep}$ were some small fraction of $R_\rL$, our models would be invalid.}
\end{center}
\end{figure}

An important aspect of the $\Omega=0$ approximation relates to the topology of the field lines.
In the models we have constructed all lines are closed (the vertical field line at the pole would formally close at infinity).
This is in contrast to the magnetic field topology of more realistic magnetosphere models with rotation; 
these have open field lines that cross the light cylinder and extend to infinity and closed lines that cross
the equator and return back to the star. The separatrix between open and closed field lines is a prominent feature of such 
models. Fig.~\ref{fig:cartoon} provides the schematic structure of this type of magnetosphere.  
As indicated in  Fig.~\ref{fig:cartoon}, the separatrix does not have to intersect the light cylinder. In terms of radial
distance in the equatorial plane these two boundaries are arranged such that $R_{\rm sep} \leq R_\rL$ 
(the separatrix cannot extend beyond $R_\rL$  without the particles moving along it violating the speed of light limit)
 
Many magnetosphere models are constructed assuming that $R_{\rm sep} = R_\rL$ exactly.
However, this choice is not dictated by some deeper physical principle. In fact, it has been suggested that $R_{\rm sep}$ 
should always sit some way inside the light cylinder \citep{uzdensky}, as in Fig.~\ref{fig:cartoon}.
Unfortunately, the local light cylinder analysis of \citet{uzdensky} does not specify the precise location of the separatrix. In the few papers presenting solutions with $R_{\rm sep} < R_\rL$, the ratio of the two radii is taken as a free parameter, and it is not immediately clear how this quantity should be contrained \citep{goodwin,timokhin06}.

If the separatrix is located close to the light cylinder, then  
$\Omega=0$ should be a good approximation for the near-surface neutron star magnetosphere because locally (i.e. out to distances $\varpi \ll R_\rL,R_{\rm sep}$) the closed field line region will dominate --- see Fig.~\ref{fig:cartoon}. In the context of magnetars, this should include the equatorial lobe with currents flowing (region I) and the large vacuum region surrounding it (II).
The only near-field domain where the  approximation is not accurate is a small region around the symmetry axis (III), which would become the largest region at radii approaching that of the separatrix. No matter 
how slow the rotation of a real system is, some field lines must be open, reaching out to the light cylinder; in this domain one should solve the full pulsar equation. Our model does not account for the presence of these open field lines, or any phenomena associated with them. Our zero-rotation approximation would be unreasonable if $R_{\rm sep} \ll R_\rL$, as the closed-field line region would shrink dramatically --- but as we have no physical reason to expect this\footnote{An argument against a large disparity between $R_{\rm sep}$ and $R_\rL$ can be based on dimensional grounds alone: the only characteristic lengthscale of the magnetosphere is $R_\rL$ and therefore one would expect $R_{\rm sep} \sim R_\rL$. An additional argument supporting this expectation can be provided by a dimensional analysis of the electromagnetic spindown torque for 
an aligned rotator. The torque formula for a neutron star with a magnetosphere like the one shown in Fig.~\ref{fig:cartoon} could be of the form
$T = T_0 (R_\rL / R_{\rm sep})^\kappa$
where $T_0$ is an amplitude comparable to the electromagnetic torque in vacuum and  $\kappa$ is unspecified. Then, unless $\kappa=0$, 
a ratio $R_\rL/R_{\rm sep}$ significantly different than unity would translate to a large deviation from the `canonical' torque $T\sim T_0$.}, 
we consider our results to be representative of the immediate exterior of a magnetar. This is nonetheless a problem that clearly 
deserves further attention.


\section*{Acknowledgments}

KG is supported by the Ram\'{o}n y Cajal Programme of the Spanish Ministerio de Ciencia e Innovaci\'{o}n,
and both he and SKL acknowledge funding from the German Science Foundation (DFG) via SFB/TR7. 
NA is supported by STFC in the UK.



\begin{thebibliography}{}


\bibitem[\protect\citeauthoryear{Beloborodov \& Thompson}{{Beloborodov \& Thompson}}{2007}]{BT07}
Beloborodov A.M., Thompson C., 2007, ApJ, 657, 967

\bibitem[\protect\citeauthoryear{Ciolfi et al.}{{Ciolfi et al.}}{2009}]{ciolfi09}
Ciolfi R., Ferrari V., Gualtieri L., Pons J. A., 2009, MNRAS, 397, 913


\bibitem[\protect\citeauthoryear{Ciolfi, Ferrari \& Gualtieri}{{Ciolfi, Ferrari \& Gualtieri}}{2010}]{ciolfi10}
Ciolfi R., Ferrari V., Gualtieri L., 2010, MNRAS, 406, 2540

\bibitem[\protect\citeauthoryear{Contopoulos et al. 1999}{{Contopoulos et al.}}{1999}]{CKF}
Contopoulos I., Kazanas D., Fendt, C.,1999, ApJ, 511, 351

\bibitem[\protect\citeauthoryear{Duncan \& Thompson}{1992}]{DT92}
Duncan R.C., Thompson C., 1992, ApJ, 392, L9


\bibitem[\protect\citeauthoryear{Goldreich \& Julian}{1969}]{GJ69}
Goldreich P., Julian W.H., 1969, ApJ, 157, 869 

\bibitem[\protect\citeauthoryear{Goodwin et al.}{2004}]{goodwin}
Goodwin S. P., Mestel J., Mestel L., Wright G.A.E., 2004, MNRAS, 349, 213 

\bibitem[\protect\citeauthoryear{Grad \& Rubin}{1958}]{grad_rubin}
Grad H., Rubin H., 1958, in Proc. 2nd Int. Conf. on Peaceful Uses of Atomic Energy. United Nations, Geneva, 31, 190 



\bibitem[\protect\citeauthoryear{Haskell et al.}{2008}]{haskell08}
Haskell, B., Samuelsson L., Glampedakis K., Andersson, N., 2008, MNRAS, 385, 531



\bibitem[\protect\citeauthoryear{Kalapotharakos et al.}{2012}]{kalapotharakos12}
Kalapotharakos C., Kazanas D., Harding A., Contopoulos I., 2012, ApJ, 749, 2

\bibitem[\protect\citeauthoryear{Lander \& Jones}{{Lander \& Jones}}{2009}]{lander09}
Lander S.K., Jones D.I., 2009, MNRAS, 395, 2162

\bibitem[\protect\citeauthoryear{Lander \& Jones}{{Lander \& Jones}}{2012}]{lander12}
Lander S.K., Jones D.I., 2012, MNRAS, 424, 482

\bibitem[\protect\citeauthoryear{Lander, Andersson \& Glampedakis}{{Lander, Andersson \& Glampedakis}}{2012}]{LAG}
Lander S.K., Andersson N., Glampedakis K., 2012, MNRAS, 419, 732


\bibitem[\protect\citeauthoryear{Mestel}{{Mestel}}{1999}]{Mestelbook}
Mestel L., 1999, Stellar Magnetism. Oxford Univ. Press, Oxford

\bibitem[\protect\citeauthoryear{Michel}{1973}]{michel73}
Michel F.C., 1973, ApJ, 180, L133

\bibitem[\protect\citeauthoryear{Scharlemann \& Wagoner}{{Scharlemann \& Wagoner}}{1973}]{sw73}
Scharlemann E.T, Wagoner R.V., 1973, ApJ, 182, 951

\bibitem[\protect\citeauthoryear{Shafranov}{1958}]{shafranov}
Shafranov V.D., 1958, Soviet J. Exp. Th. Phys, 6, 545

\bibitem[\protect\citeauthoryear{Shapiro \& Teukolsky}{{Shapiro \& Teukolsky}}{1983}]{STbook}
Shapiro S.L., Teukolsky S.A., 1983, Black Holes, White Dwarfs and Neutron Stars: 
The Physics of Compact Objects. Wiley, New York


\bibitem[\protect\citeauthoryear{Spitkovsky}{{Spitkovsky}}{2006}]{spitkovsky06}
Spitkovsky A., 2006, ApJ, 648, L51


\bibitem[\protect\citeauthoryear{Thompson et al.}{2002}]{thompson02}
Thompson C., Lyutikov M., Kulkarni S.R., 2002, ApJ, 574, 332

\bibitem[\protect\citeauthoryear{Timokhin}{2006}]{timokhin06}
Timokhin A. N., 2006, MNRAS, 368, 1055

\bibitem[\protect\citeauthoryear{Tomimura \& Eriguchi}{{Tomimura \& Eriguchi}}{2005}]{tomi_eri}
Tomimura Y., Eriguchi Y., 2005, MNRAS, 359, 1117


\bibitem[\protect\citeauthoryear{Parfrey et al.}{{Parfrey et al.}}{2012}]{parfrey12}
Parfrey K., Beloborodov A.M., Hiu L., 2012, ApJ, 754, L12

\bibitem[\protect\citeauthoryear{Pavan et al.}{{Pavan et al.}}{2009}]{pavan09}
Pavan L., Turolla R., Zane S., Nobili L., 2009, MNRAS, 395, 753


\bibitem[\protect\citeauthoryear{Vigan\`o et al.}{{Vigan\`o et al.}}{2011}]{vigano11}
Vigan\`o D., Pons J.A., Miralles J.A., 2011, A\&A, 533, A125


\bibitem[\protect\citeauthoryear{Uzdensky}{{Uzdensky}}{2003}]{uzdensky}
Uzdensky D.A., 2003, ApJ, 598, 446


\end{thebibliography}
\end{document}